\documentclass[twocolumn]{aastex62}
\received{}
\revised{}
\accepted{}
\submitjournal{ApJL, Accepted on September 20, 2019}

\usepackage{placeins}
\usepackage{amsmath}
\usepackage{color}
\usepackage{enumerate}

\begin{document}

\newcommand\msun{$M_\odot$}
\newcommand\lsun{$L_\odot$}
\newcommand\gcmtwo{${\rm g~cm}^{-2}$}
\newcommand\gcmthree{${\rm g~cm}^{-3}$}
\newcommand\be{\begin{equation}}
\newcommand\en{\end{equation}}
\newcommand{\rev}[1]{\textcolor{red}{#1}}
\newcommand{\add}[1]{\textcolor{blue}{#1}}

\shorttitle{AN IDEAL TESTBED FOR PLANET-DISK INTERACTION: PDS~70} 
\shortauthors{Bae et al.}

\title{AN IDEAL TESTBED FOR PLANET-DISK INTERACTION: TWO GIANT PROTOPLANETS IN RESONANCE SHAPING THE PDS~70 PROTOPLANETARY DISK}

\correspondingauthor{Jaehan Bae}
\email{jbae@carnegiescience.edu}

\author[0000-0001-7258-770X]{Jaehan Bae}
\affil{Department of Terrestrial Magnetism, Carnegie Institution for Science, 5241 Broad Branch Road NW, Washington, DC 20015, USA}
\affil{NHFP Sagan Fellow}

\author[0000-0003-3616-6822]{Zhaohuan Zhu}
\affil{Department of Physics and Astronomy, University of Nevada, Las Vegas, 4505 South Maryland Parkway, Las Vegas, NV 89154, USA}

\author{Cl{\'e}ment Baruteau}
\affil{IRAP, Universit{\'e} de Toulouse, CNRS, UPS, F-31400 Toulouse, France}

\author[0000-0002-7695-7605]{Myriam Benisty}
\affiliation{Departamento de Astronom\'ia, Universidad de Chile, Camino El Observatorio 1515, Las Condes, Santiago, Chile}
\affiliation{Unidad Mixta Internacional Franco-Chilena de Astronom\'ia, CNRS, UMI 3386}
\affiliation{Univ. Grenoble Alpes, CNRS, IPAG, 38000 Grenoble, France.}

\author[0000-0002-7078-5910]{Cornelis P.~Dullemond}
\affiliation{Zentrum f\"ur Astronomie, Heidelberg University, Albert Ueberle Str.~2, 69120 Heidelberg, Germany}

\author[0000-0003-4689-2684]{Stefano Facchini}
\affiliation{European Southern Observatory, Karl-Schwarzschild-Str. 2, 85748 Garching, Germany}

\author[0000-0002-0786-7307]{Andrea Isella}
\affiliation{Department of Physics and Astronomy, Rice University, 6100 Main Street, MS-108, Houston, TX 77005, USA}

\author[0000-0001-7250-074X]{Miriam Keppler}
\affiliation{Max Planck Institute for Astronomy, K\"onigstuhl 17, 69117, Heidelberg, Germany}

\author[0000-0002-1199-9564]{Laura M. P\'erez}
\affiliation{Departamento de Astronom\'ia, Universidad de Chile, Camino El Observatorio 1515, Las Condes, Santiago, Chile}

\author[0000-0002-0786-7307]{Richard Teague}
\affil{Department of Astronomy, University of Michigan, 311 West Hall, 1085 S. University Ave, Ann Arbor, MI 48109, USA}
\affil{Harvard-Smithsonian Center for Astrophysics, 60 Garden Street, Cambridge, MA 02138, USA}

\begin{abstract}

While numerical simulations have been playing a key role in the studies of planet-disk interaction, testing numerical results against observations has been limited so far. With the two directly imaged protoplanets embedded in its circumstellar disk, PDS~70 offers an ideal testbed for planet-disk interaction studies. Using two-dimensional hydrodynamic simulations we show that the observed features can be well explained with the two planets in formation, providing strong evidence that previously proposed theories of planet-disk interaction are in action, including resonant migration, particle trapping, size segregation, and filtration. Our simulations suggest that the two planets are likely in 2:1 mean motion resonance and can remain dynamically stable over million-year timescales. The growth of the planets at $10^{-8}-10^{-7}~M_{\rm Jup}~{\rm yr}^{-1}$, rates comparable to the estimates from H$\alpha$ observations, does not destabilize the resonant configuration. Large grains are filtered at the gap edge and only small, (sub-)$\mu$m grains can flow to the circumplanetary disks and the inner circumstellar disk. With the sub-millimeter continuum ring observed outward of the two directly imaged planets, PDS~70 provides the first observational evidence of particle filtration by gap-opening planets.  The observed sub-millimeter continuum emission at the vicinity of the planets can be reproduced when (sub-)$\mu$m grains survive over multiple circumplanetary disk gas viscous timescales and accumulate therein. One such possibility is if (sub-)$\mu$m grains grow in size and remain trapped in pressure bumps, similar to what we find happening in circumstellar disks. We discuss potential implications to planet formation in the solar system and mature extrasolar planetary systems.

\end{abstract}


\section{INTRODUCTION}
\label{sec:introduction}

Numerical simulations have been playing a key role in the studies of planet-disk interaction and planet formation, by confirming analytic theories, examining non-linear phenomenons, allowing us to explore a broad parameter space, and making observational predictions. Despite the important role, testing results from numerical simulations against observations has been limited because observing planets in formation has been very challenging. The situation is however gradually changing. Thanks to increasingly powerful observing facilities and techniques we are now able to peer into the birthplaces of planets, routinely finding disk substructures hinting at on-going planet formation \citep[e.g.,][]{avenhaus18,andrews18,pinte18,Teague18}, although whether or not planets are indeed the cause of the observed features has to be further investigated.

With two directly detected protoplanets embedded in its circumstellar disk \citep{keppler18,wagner18,haffert19}, PDS~70 offers an ideal testbed for planet-disk interaction studies. The two planets, PDS~70b and c, are located 195 and 234~milliarcsecond (hereafter mas; \citealt{keppler18,haffert19}) from the $5.4$~Myr-old, K7 pre-main sequence star PDS~70 \citep{muller18} on the sky. The de-projected distances between the planets and the star ($\sim20$ and 35~au) suggest that the two planets are in or near 2:1 mean motion resonance \citep{haffert19}. They cleared their vicinity and opened an inner cavity in the circumstellar disk, initially identified in IR observations \citep{hashimoto12,dong12}. Sub-mm continuum observations have also revealed an inner cavity with the emission concentrated beyond the two planets' orbits \citep{hashimoto15,long18,keppler19}. Rotation velocity measurements of CO gas revealed that the gas pressure changes over radius in a way suggesting that grains are trapped in a pressure bump \citep{keppler19}, although it is pointed out that the cavity is too wide to explain with PDS~70b alone, the only planet known to be in the system at that time. The size of the inner cavity is larger in sub-millimeter (hereafter sub-mm) continuum ($\sim74$~au; \citealt{keppler19}) than in IR ($\sim52$~au with the Gaia DR2 distance of 113~pc; \citealt{dong12}). IR and sub-mm observations suggest that both planets are surrounded by dusty circumplanetary disk \citep[CPD;][]{christiaens19,isella19}, while the nature of the sub-mm emission detected near PDS~70b, which is 74~mas offset from the location of PDS~70b inferred from H$\alpha$/IR emission, is yet to be explained \citep{isella19}.

Interestingly enough, many of the aforementioned observed features have been previously seen in and proposed to happen in protoplanetary disks by numerical simulations. Multiple planets in resonance are proposed as a possible cause of inner cavities seen in (pre-)transitional disks \citep{zhu11}. In the pressure bump forming beyond a planet's orbit, particles can be efficiently trapped and we expect segregation of grain sizes such that, under typical protoplanetary disk conditions, smaller grains are distributed over a larger radial extent, resulting in a smaller cavity size \citep{pinilla12b,zhu12}. The outer edge of the gap can act as a {\it filter} so small particles can penetrate the planet-induced gap whereas large particles are filtered and remain trapped beyond the gap \citep{rice06,zhu12}. The existence of circumplanetary disks around giant planets is also conceptually and numerically predicted \citep{quillen98,lubow99,ayliffe09,ward10}, as a consequence of angular momentum conservation similar to the formation of circumstellar disks around protostars. 

In this paper, we carry out two-dimensional hydrodynamic calculations adopting physical properties of the PDS~70 circumstellar disk and the two planets embedded therein. By simulating the dynamics of disk gas and dust in response to two growing planets, we show that the  signposts of planet-disk interaction predicted by numerical simulations and the observed features of the PDS~70 disk show a good agreement with each other. We believe the resemblance between simulations and observations strongly supports the previously proposed theories, including resonant migration, particle trapping, size segregation, and filtration, are in action.

The paper is organized as follows. In Section \ref{sec:setup}, we introduce our numerical model. In Section \ref{sec:results}, we present results from numerical simulations, focusing on the evolution of the circumstellar disk and planets' orbits. In Section \ref{sec:discussion}, we generate simulated sub-mm continuum images from our simulations to compare with the observations and discuss how our findings can help better understand planet-disk interaction, CPDs, and planet formation. We present our conclusions in Section \ref{sec:conclusion}.

\section{MODEL SETUP}
\label{sec:setup}

\subsection{Initial Gas Disk}

We adopt an initial disk gas surface density profile that falls off with an exponential tail against radius $R$
\be
\label{eqn:sigma}
\Sigma_{\rm gas, init}(R)   =  \Sigma_c \left({R \over  R_c}\right)^{-1} \exp{ \left(- {R \over R_c} \right)},
\en
where we choose $R_c=40$~au and $\Sigma_c=2.7~{\rm g~cm}^{-2}$ so that the total disk gas mass within the simulation domain is 0.003~\msun, similar to the model used in \citet{keppler19}.

To set up the initial disk temperature profile, we iterate Monte Carlo radiative transfer (MCRT) calculations using RADMC-3D \citep{radmc3d} until we obtain a converged three-dimensional disk density and temperature profile (see Appendix \ref{sec:disk_temperature}).
The least-squares power-law fit to the density-weighted, vertically integrated temperature $\bar{T}$ is 
\be
\label{eqn:temperature}
\bar{T}(R) = 44~{\rm K} \left( {R \over 22~{\rm au}} \right)^{-0.24}.
\en
Adopting a stellar mass of $0.85~M_\odot$ \citep{keppler19} and a mean molecular weight of 2.4, this temperature profile corresponds to a disk aspect ratio $H/R$ of 
\be
H/R = 0.067 \left( {R \over 22~{\rm au}} \right)^{0.38}.
\en

\subsection{Planets}

We fix PDS 70b's mass to 5 $M_{\rm Jup}$, consistent with previous estimates  \citep{keppler18,muller18}. We test three different masses for PDS 70c: 2.5, 5, and 10 $M_{\rm Jup}$. The two planets are initially placed at 20 and 35~au, in an agreement with the observed de-projected distances to the central star \citep{keppler19,haffert19}. With the semi-major axes the orbital period ratio is 2.3:1, so the two planets are located slightly outside of 2:1 mean motion resonance initially. In addition to the three simulations, we carry out a simulation with PDS~70b only. 

We run simulations for 2~Myr, which corresponds to about 20,000 orbits at PDS 70b's initial radial location. We linearly increase planet masses over the first $10^4$ years, while fixing their orbits. After $10^4$ years, we allow the planets to gravitationally interact with each other as well as with the circumstellar disk. As we will show below, the two planets settle into 2:1 mean motion resonance within about 0.1 Myr in all three cases and the common gap opened by the planets reaches a quasi-steady state by $0.2$~Myr.

To examine whether increases in planet masses affect the stability of the system, we allow the planets to accrete gas starting at $0.2$~Myr. In practice, the gas density in cells within a fraction of planets' Hill radius is reduced by a fraction $f_{\rm acc} \Omega \Delta t$ and added to the planets each hydrodynamic time step, following the approach presented in \citet{kley99} and \citet{durmann17}. This results in the gas depletion timescale of $\tau_{\rm acc} = (f_{\rm acc} \Omega)^{-1}$. We choose $f_{\rm acc} = 0.01$, with which the planets accrete at $10^{-8} - 10^{-7}~M_{\rm Jup}~{\rm yr}^{-1}$, consistent with the estimates based on H$\alpha$ observations \citep{wagner18,haffert19}. Note that this parameterized planet accretion is adopted to examine the dynamical stability in a way that the total mass and momentum are conserved, rather than to realize actual accretion processes within planets' Hill sphere.

The gravitational potential of planets is softened over $60\%$ of the local gas scale height, in order to mimic the overall magnitude of the torque in three dimensions \citep{muller12}.

\subsection{Hydrodynamic Simulations}
\label{sec:methods_hydro}

We carry out two-dimensional, locally isothermal hydrodynamical simulations using the Dusty FARGO-ADSG code \citep{baruteau19}. This is an extended version of the publicly available FARGO-ADSG \citep{masset00,baruteau08a,baruteau08b}, with Lagrangian test particles implemented \citep{baruteau16}.

The simulation domain extends from 2.2 to 198~au in the radial direction and covers the entire $2\pi$ in azimuth. We adopt 672 logarithmically-spaced grid cells in the radial direction and 936 uniformly-spaced grid cells in the azimuthal direction. At the radial boundaries we adopt a wave-damping zone \citep{devalborro06} to suppress wave reflection, from 2.2 to 2.64~au and from 176 to 198~au. Because we are interested in long-term, Myr-timescale evolution, we decrease the surface density in the wave-damping zone $\Sigma_{\rm gas, damp}$ over the local viscous timescale following
\begin{equation}
\Sigma_{\rm gas, damp} = \Sigma_{\rm gas, init} \exp{(-t/t_{\rm vis})}.
\end{equation}
Here, $t_{\rm vis} = R^2/\nu$ is the viscous timescale and $\nu = \alpha c_s^2 /\Omega$, where $\alpha$ is the viscosity parameter, $c_s$ is the sound speed, and $\Omega$ is the orbital frequency.
A uniform disk viscosity $\alpha=10^{-3}$ is applied.

In addition to the gas component we insert $100,000$ Lagrangian test particles at $t=0.5$~Myr, well after the overall disk structure reaches a quasi-steady state. Test particles are inserted between 50 and 100~au, with a uniform dust-to-gas mass ratio of $2.5~\%$ across this radial region. This results in a total dust mass of about $10~M_\earth$. We assume a dust bulk density of $1.26~{\rm g~cm}^{-3}$, which corresponds to that of aggregates with $30~\%$ silicate matrix and $70~\%$ water ice. Particle sizes are determined such that we have approximately same number of test particles per decade of size between 0.1~$\mu$m and 1~mm. We choose the maximum particle size of 1~mm because the fragmentation-limited maximum grain size \citep[e.g.,][]{birnstiel12} between 50 and 100~au is a few hundred $\mu$m with the disk and particle properties we adopt. With the initial gas surface density in Equation (\ref{eqn:sigma}), the Stokes number of the test particles can be expressed as 
\begin{equation}
{\rm St} \simeq 0.07 \left( {s \over {1~{\rm mm}}} \right) \left( {R \over {40~{\rm au}}} \right) \exp \left( {R \over {40~{\rm au}}} \right).
\end{equation}

Test particles feel the gravity of the star and the planets. In addition, they interact with the circumstellar disk gas via aerodynamic drag. Turbulent diffusion is included as stochastic kicks on the particles' position following the method presented in \citet{charnoz11}, adopting $\alpha=10^{-3}$. Test particles do not provide feedback onto the planets and the disk gas. The size evolution of particles is not included in the simulations. 

Since we aim to explain the sub-mm continuum flux associated with the CPDs (Section \ref{sec:cpd}), we assign mass to test particles so that we can keep track of the CPD dust mass. In practice, this mass assignment is done such that the dust mass at each radius in the initial disk is distributed over a range of dust size $s$, from $0.1~\mu$m to 1~mm, to have the mass per interval in $\log(s)$ be proportional to $s^{0.5}$ (this corresponds to a dust size distribution of $n(s) \propto s^{-3.5}$).

\section{RESULTS}
\label{sec:results}

\subsection{Disk Evolution}

\begin{figure*}[th]
    \centering
    \includegraphics[width=0.92\textwidth]{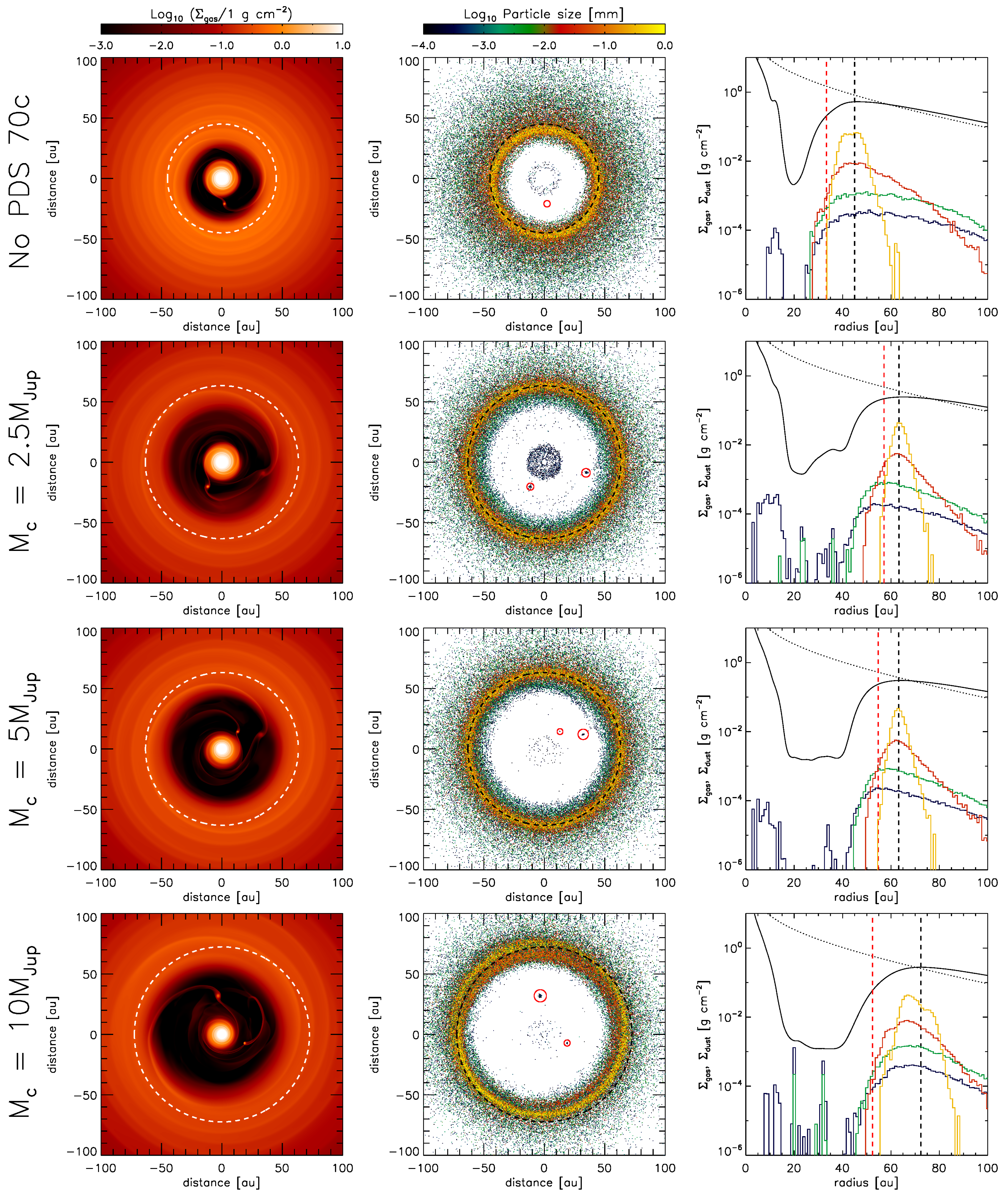}
    \caption{(Left) The two-dimensional gas surface density distribution, (middle) particle distribution, and (right) azimuthally-averaged radial gas and dust density distributions at $t=0.6$~Myr. From top to bottom, models without PDS~70c, with $M_c = 2.5~M_{\rm Jup}$, $M_c = 5~M_{\rm Jup}$, and $M_c = 10~M_{\rm Jup}$. Dashed circles in the left and middle panels show the location of the azimuthally-averaged gas pressure maximum. In the middle panels, the red circles show the size of Hill radius of the planets. In the right panel, the red dashed line shows 1:2 outer Lindblad resonance location of PDS~70c (PDS~70b in the model without PDS~70c), while the black dashed line show the gas pressure maximum. The dotted curve shows the initial gas surface density profile. The color histograms show the azimuthally-averaged dust surface density with sizes of (dark blue) $0.1-1~\mu$m, (green) $1-10~\mu$m, (red) $10-100~\mu$m, and (yellow) $0.1-1$~mm, adopting 1~au bin size in radius.}
    \label{fig:fig1}
\end{figure*}

We start by discussing the overall circumstellar disk evolution. In Figure \ref{fig:fig1}, we present the two-dimensional gas surface density distribution, test particle distribution, and azimuthally-averaged radial distributions of the gas and dust surface density at $t=0.6$~Myr. When only PDS~70b exists in the disk, the planet's outer gap edge becomes eccentric ($e \sim 0.1$) because of the eccentric Lindblad resonance \citep{lubow91a,lubow91b,papaloizou01}. The wave modes from the circular component of the planet's potential are excited at $R_{\rm LR, circ}= [(m\pm1)/m]^{2/3}~R_p$, where $m$ denotes the azimuthal wavenumber of wave modes. The outer Lindblad resonance located farthest away from the planet is the $m=1$ mode, which occurs at $1.6~R_p \simeq 32$~au. As shown in the radial gas density profile in Figure \ref{fig:fig1}, PDS~70b opens a wide gap with the gas density peaking at about 45~au. The total angular momentum exchange via the circular component of the planet's potential, which is the summation of the individual contribution over the entire azimuthal wavenumbers, is therefore significantly reduced. On the other hand, wave modes from the eccentric component of the planet's potential launch at $R_{\rm LR, ecc}= [(m \pm l)/m]^{2/3}~R_p$ ($l$ is an integer greater than 1), which is always beyond the Lindblad resonance of their counterpart circular component $R_{\rm LR, circ}$ in the outer disk. Thus, the overall angular momentum exchange can be dominated by the eccentric component, making the outer gap edge eccentric. In the PDS~70b-only model particles with sizes $0.1-1$~mm (hereafter sub-mm particles), which dominate the sub-mm continuum flux, are trapped in the gas pressure peak at 45~au; this is insufficient to explain the continuum peak at $\sim74$~au in the sub-mm observation \citep{keppler19}. 

\citet{muley19} recently showed that accreting planets can have an abrupt increase in its orbital eccentricity to $e \sim 0.25$ as they grow in mass. These planets can carve a wider gap than otherwise, helping explain the large continuum cavity. We however do not observe such an increase in orbital eccentricity, at least for the duration of our simulations. We conjecture this could be because the reported eccentricity growth prefers large accretion rates. In the fiducial simulation of \citet{muley19}, the time-averaged accretion rate until the onset of the eccentricity growth is $\sim 2.5~M_{\rm Jup}/3.5~{\rm Myr} \simeq 7 \times 10^{-7}~M_{\rm Jup}~{\rm yr}^{-1}$, more than an order of magnitude larger than the  accretion rate estimates from observations.

When both PDS~70b and c are inserted the two planets open a common gap. The gas disk and particle ring at the common gap outer edge remain circular when PDS~70c's mass is 2.5 or 5~$M_{\rm Jup}$. In the two models, sub-mm particles are trapped in the pressure peak at $\sim63$~au, which is further out compared with the PDS~70b-only model and is in a better agreement with the continuum ring location in the sub-mm observation \citep{keppler19}. The gas surface density around PDS~70c is larger than that around PDS~70b when $M_c = 2.5~M_{\rm Jup}$, while it is comparable to the gas density around PDS~70b when $M_c = 5~M_{\rm Jup}$. This, together with the fact that the CO emission is significantly more depleted around PDS~70b's orbit \citep{keppler19}, suggests that PDS~70c likely has a smaller mass than PDS~70b.

When PDS~70c's mass is 10~$M_{\rm Jup}$ the outer gap edge in the gas disk and particle ring become eccentric, with the eccentricity varying between 0.2 and 0.4 over time. Similar to the PDS~70b-only model, the non-zero disk eccentricity arises as the gap is sufficiently wide such that the eccentric component of PDS~70c's potential dominates over its circular component. Because such a large continnuum ring eccentricity of $e \sim 0.2-0.4$ can be ruled out by sub-mm observations \citep{keppler19}, we conclude that PDS~70c's mass has to be smaller than 10~$M_{\rm Jup}$.

\begin{figure}[h!]
    \centering
    \includegraphics[height=0.85\textheight]{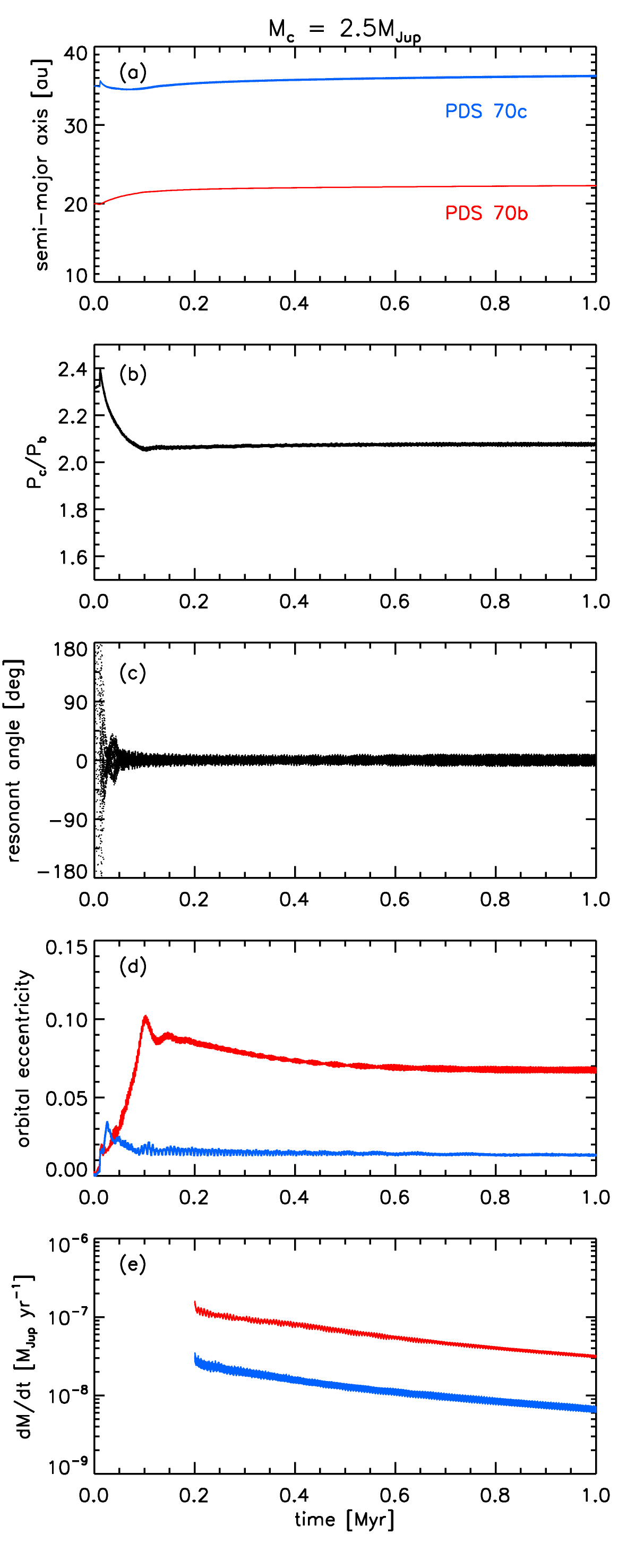}
    \caption{Time evolution of (a) semi-major axes, (b) orbital period ratio, (c) resonant angle, (d) orbital eccentricities, and (e) accretion rates onto the planets. Note that the two planets settle into 2:1 mean motion resonance within the first 0.1~Myr and remain dynamically stable for the duration of the simulation. The first 1~Myr of the simulations are shown for visualization purposes only.}
    \label{fig:orbit1}
\end{figure}

\begin{figure*}[h!]
\plottwo{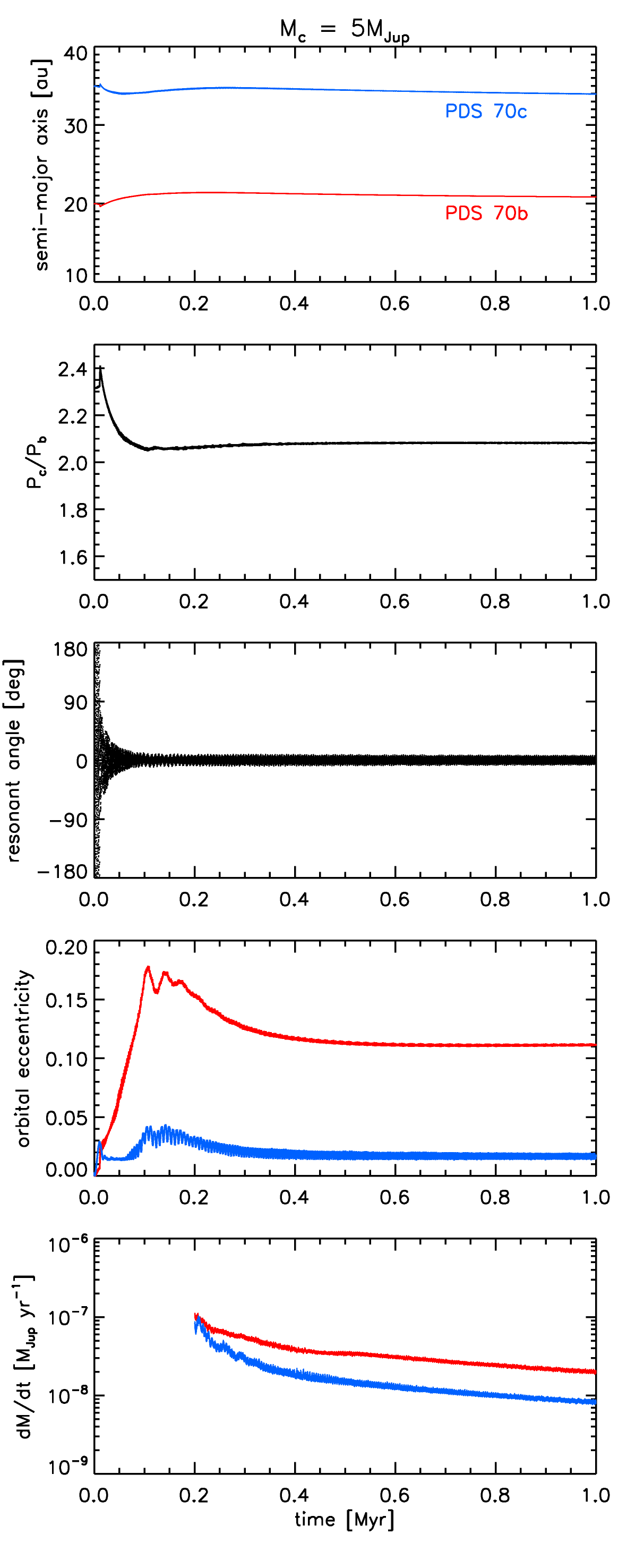}{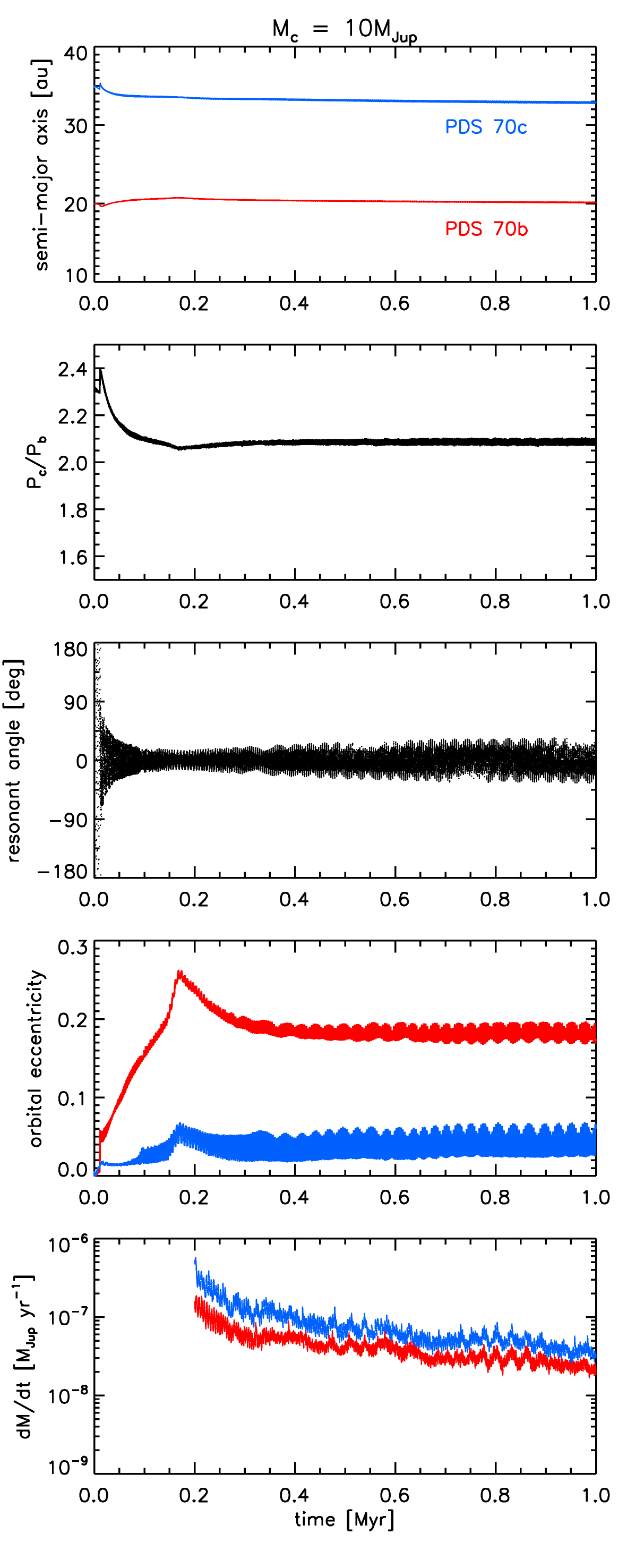}
\caption{Same as Figure \ref{fig:orbit1}, but results with (left panels) $M_c = 5~M_{\rm Jup}$ and (right panels) $M_c = 10~M_{\rm Jup}$.}
\label{fig:orbit2}
\end{figure*}

\subsection{Planets' Orbital Evolution}

We plot orbital elements of the two planets in Figure \ref{fig:orbit1} and \ref{fig:orbit2}. As can be seen from the resonant angle, defined as $\psi_{cb} = 2 \lambda_b - \lambda_c - \varpi_c$ where $\lambda_b$ and $\lambda_c$ are mean longitudes of the planets, and $\varpi_c$ is the outer planet's longitude of perihelion, the two planets migrate toward each other and settle into 2:1 mean motion resonance within the first 0.1~Myr of the simulations for all three cases. We find that the two planets remain dynamically stable for the next 2~Myr.    Such a rapid adjustment in their orbits into a resonant configuration suggests that PDS~70b and c are likely in 2:1 mean motion resonance. The exact period ratio is slightly larger than 2:1 because the planets experience repulsion as they interact with each other's spiral arms \citep{baruteau13}. 

\begin{figure*}[th]
    \centering
    \includegraphics[width=1\textwidth]{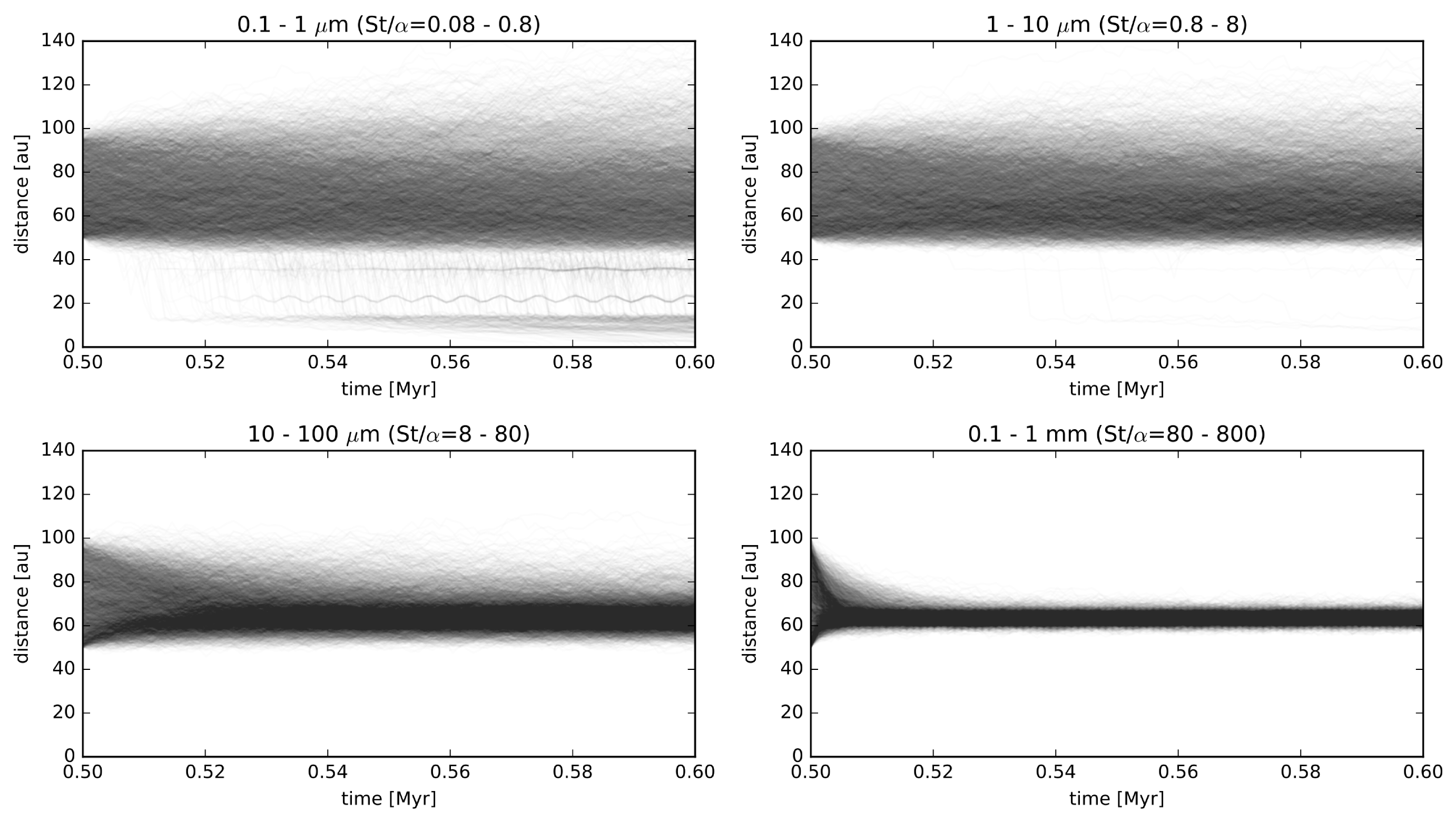}
    \caption{An overlay of the trajectories of test particles in $M_c = 2.5~M_{\rm Jup}$ model: (top left) $0.1-1~\mu$m, (top right) $1-10~\mu$m, (bottom left) $10-100~\mu$m, and (bottom right) $0.1-1$~mm particles. At the gas pressure peak (63~au) particles have Stokes numbers St $\simeq 0.8~(s/1~{\rm mm})$. Note that only small particles with sizes $\lesssim 1~\mu$m penetrate the common gap opened by the planets. Note also that some of the small particles penetrating the gap are captured in the circumplanetary disks.}
    \label{fig:partevol}
\end{figure*}

When PDS~70c's mass is $2.5~M_{\rm Jup}$ the two planets migrate outward (Figure \ref{fig:orbit1}a). Between 0.2 and 2~Myr, PDS~70b and c's semi-major axes increase at 0.30  and 0.51~au~Myr$^{-1}$, respectively. On the other hand, when PDS~70c's mass is comparable to or larger than PDS~70b's mass ($M_c=5~M_{\rm Jup}$ and $10~M_{\rm Jup}$ models), the planets experience an inward migration (Figure \ref{fig:orbit2}). When  PDS~70c's mass is $5~M_{\rm Jup}$, PDS~70b and c's semi-major axes decrease at 0.39 and 0.73~au~Myr$^{-1}$. When PDS~70c's mass is $10~M_{\rm Jup}$, PDS~70b and c's semi-major axes decrease at  0.27  and 0.43~au~Myr$^{-1}$, respectively.

In all three models PDS~70c's orbital eccentricity remains relatively small, less than about $0.05$, whereas PDS~70b's orbital eccentricity converges to $e \sim 0.05 - 0.2$ with the exact value dependent upon PDS~70c's mass. Note that a smaller orbital eccentricity for the outer planet is known to be a generic feature of 2:1 mean motion resonance \citep{baruteau13}. Even the largest PDS~70b's orbital eccentricity we find in our simulations ($e \sim 0.2$) cannot be ruled out with the existing observations \citep{muller18}. 

The accretion rates onto the planets gradually decrease over time. Our simulations suggest that the mass growth at rates consistent with recent H$\alpha$ observations ($10^{-8} - 10^{-7}~M_{\rm Jup}~{\rm yr}^{-1}$; \citealt{wagner18,haffert19}) is unlikely to destablize the mean motion resonance.

\section{DISCUSSION}
\label{sec:discussion}

\subsection{Particle Trapping, Filtration, and Size Segregation}

Giant planets open a deep gap and trap large grains at the gap edge \citep[e.g.,][]{Paardekooper+04,Fouchet2007,zhu12}. The ratio between the Stokes number and the disk viscosity parameter St/$\alpha$ is important in determining the width of particle distribution in a pressure bump and whether or not particles penetrate the gap \citep{zhu12,dullemond18}: large particles having Stokes number St $\gtrsim \alpha$ are efficiently filtered at the gap edge, whereas small particles with St $\lesssim \alpha$ are mixed well with gas and follow the gas distribution. 

In Figure \ref{fig:partevol} we show the trajectories of test particles from the $M_c=2.5~M_{\rm Jup}$ model. As shown, small particles with sizes $\lesssim 10~\mu$m have a broad radial distribution, whereas large particles with sizes $\gtrsim 10~\mu$m move toward the pressure bump aided by aerodynamic drag and remain trapped within a narrower radial region. While large grains incapable of penetrating the gap establish a quasi-steady state radial distribution rapidly, note that small, sub-$\mu$m grains\footnote{The exact size of grains available to penetrate a gap can differ in other disks depending upon various disk properties, including the gas surface density, the gap depth, and the level of disk turbulence, as well as the grain internal density \citep{zhu12}.} are well coupled with gas so leak gradually over time, flowing into the inner disk. It is also interesting to note that some sub-$\mu$m grains are captured in circumplanetary disks while penetrating the gap (see also Figure \ref{fig:fig1}). The rate at which small particles penetrate the gap appears to be dependent on PDS~70c's mass. Among the three models with two planets, we find that the inner disk is fed with small dust most efficiently when $M_c = 2.5~M_{\rm Jup}$. We conjecture this is because the outward resonant migration facilitates the interaction between PDS~70c and the dust reservoir in the outer disk.

\begin{figure*}[th]
    \centering
    \includegraphics[width=1\textwidth]{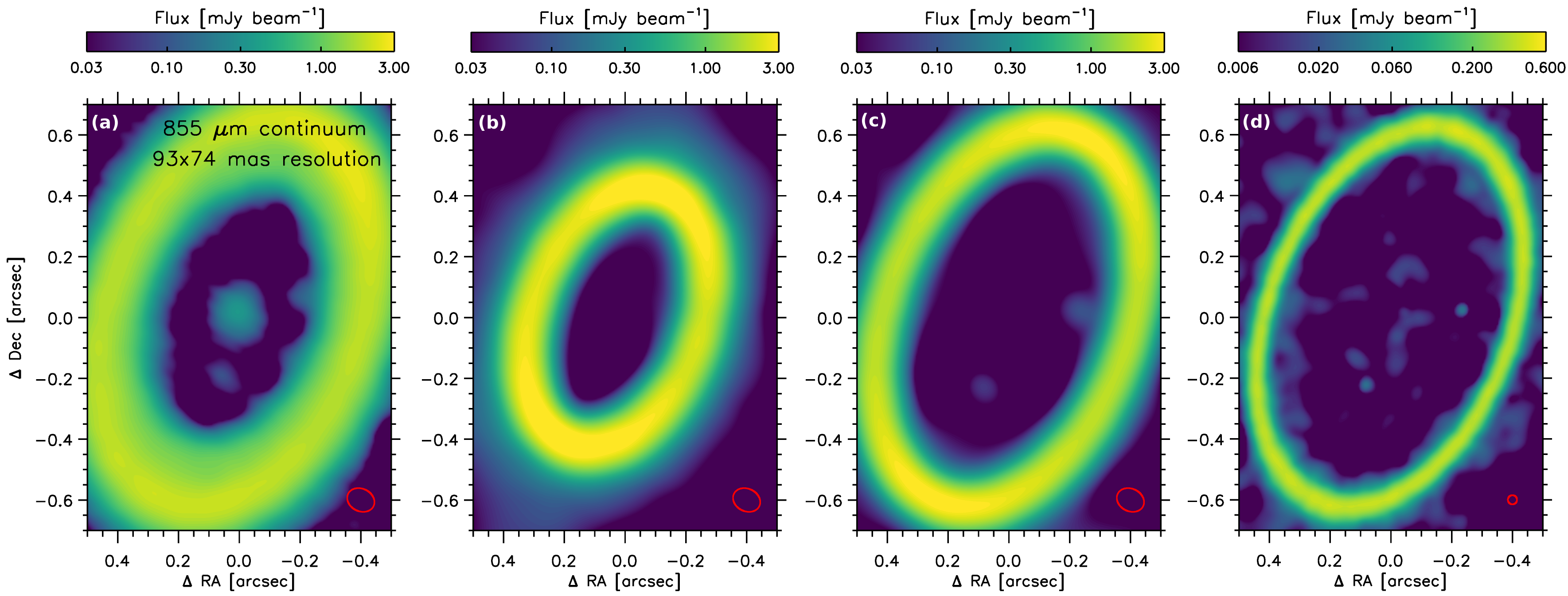}
    \caption{(a) The observed continuum emission at $855~\mu$m, reproduced from \citet{isella19}. (b) A simulated continuum image with PDS~70b-only model, using spatial resolution ($93\times74$~mas synthesized beam), rms noise level ($18~\mu{\rm Jy}~{\rm beam}^{-1}$), disk inclination ($51.7^\circ$), and position angle ($156.7^\circ$) comparable to the observation. (c) Same as middle left panel, but with $M_c=2.5~M_{\rm Jup}$ model. The CPD dust mass is distributed over a range of dust sizes from $0.1~\mu$m to 1~mm (see text). (d) Same as panel (c) but using a $30\times30$~mas synthesized beam and $14~\mu{\rm Jy}~{\rm beam}^{-1}$ rms noise, requiring 5 hours of on source integration. The synthesized beam is shown in red at the bottom right corner of each panel.}
    \label{fig:simobs}
\end{figure*}

Figure \ref{fig:simobs}a, b, and c show the observed continuum image at $855~\mu$m and simulated continuum images based on the particle distribution of PDS~70b-only model and $M_c = 2.5~M_{\rm Jup}$ model presented in Figure \ref{fig:fig1} (see Appendix \ref{sec:simobs} for details about simulated observations). As shown, $M_c = 2.5~M_{\rm Jup}$ model reproduces the flux and morphology of the outer continuum ring reasonably well, but the continuum ring in PDS~70b-only model locates closer to the center of the system compared with the observation. With the sub-mm continuum ring observed outward of the two directly imaged planets, PDS~70 system provides the first observational evidence that gap-opening planets can trap particles beyond their orbits.

Observations of transitional disks at different wavelengths, probing different regions in the disk and/or grains with different sizes, show varying cavity size. In general, the inner cavity seen in molecular gas lines and optical/IR scattered light observations is smaller in size than seen in (sub-)millimeter continuum observations \citep[e.g.,][]{vandermarel2016}. This is consistent with what is seen in PDS~70 observations \citep{hashimoto12,dong12,keppler19} and the gas and particle distribution seen in our simulations (Figure \ref{fig:fig1} and \ref{fig:partevol}).

We note that, similar to PDS~70, two planet candidates are detected within the inner cavity of the transitional disk around HD~100546. Near IR spectroscopic monitoring of fundamental ro-vibrational CO emission lines revealed a spectroastrometric evidence of an orbiting companion within the inner cavity, at a de-projected separation of $\sim12$~au from the central star \citep{brittain13,brittain14,brittain19}. In addition, a point source sub-mm continuum is detected with ALMA at a de-projected separation of 7.8~au \citep{perez19}. The de-projected distances suggest that the two planets may be in/near 2:1 mean motion resonance (orbital period ratio $\sim 1.9:1$). If confirmed, HD~100546 would offer another convincing example of giant planets in resonance opening inner cavity in transitional disk.

\subsection{Circumplanetary Disks: Can We Explain the Sub-millimeter Continuum Emission?}
\label{sec:cpd}

Sub-millimeter observations with ALMA revealed spatially unresolved continuum emission at the vicinity of PDS~70b and c, possibly originating from dusty CPDs (\citealt{isella19}; Figure \ref{fig:simobs}a). Here, we discuss if the observed continuum flux can be explained with thermal emission from the dust in the CPDs.

If the CPDs are in a steady-state such that the CPD gas density remains constant over time, we can assume that gas is supplied from the circumstellar disk onto CPDs at the rate comparable to the planets' accretion rate, $\dot{M}_{\rm acc} \simeq 10^{-8} - 10^{-7}~M_{\rm Jup}~{\rm yr}^{-1}$ \citep{wagner18,haffert19}. Coupled with the gas flow from the circumstellar disk, only small, sub-$\mu$m dust is replenished as can be seen from Figure \ref{fig:fig1} and \ref{fig:partevol}. We use $f_{\rm dtg}$ for the mass ratio between these small dust to circumstellar disk gas at the outer edge of the gap. In our simulations, we find that this is of order of $10^{-3}$ (Figure \ref{fig:fig1}). Since sub-$\mu$m grains are expected to be well coupled to the CPD gas, they would accrete onto the planets over the CPD gas viscous timescale which can be written as $\tau_{\rm vis} \simeq T_p/(5\pi\alpha_{\rm CPD})$, where $T_p$ is the planet's orbital period and $\alpha_{\rm CPD}$ is the viscosity parameter of the CPD \citep{zhu11}. We thus expect that the CPDs would have a steady-state dust mass of 
\begin{eqnarray}
\nonumber
M_{\rm dust, CPD} & \simeq & \dot{M}_{\rm acc} \tau_{\rm vis} \\
\nonumber 
& = & 2\times 10^{-5} M_\earth \left( { \dot{M}_{\rm acc} \over {10^{-8}~M_{\rm Jup}~{\rm yr}^{-1}}} \right) \left( {f_{\rm dtg} \over 10^{-3}} \right) \\
&  & \times \left( {\alpha_{\rm CPD} \over 10^{-3}} \right)^{-1} \left( { T_p \over 100~{\rm yrs}} \right).
\end{eqnarray}
Assuming semi-major axes of 22 and 35~au for PDS~70b and c and the stellar mass of $0.85~M_\odot$, the orbital periods of the planets are 112 and 225~years, respectively. This results in $M_{\rm dust, CPD}$ of  $2.2\times10^{-5}~M_\earth$ and $4.5\times10^{-5}~M_\earth$.

Adopting the distance $d=113$~pc \citep{gaia16,gaia18}, the optically thin continuum flux at $855~\mu$m (350~GHz) is
\begin{eqnarray}
\label{eqn:cont_flux}
\nonumber
F_\nu & =  &{\kappa_\nu B_\nu (T_{\rm dust, CPD}) \over d^2} M_{\rm dust, CPD} \\
&=& 0.15~\mu{\rm Jy} \left( {\kappa_\nu \over 2~{\rm cm}^2~{\rm g}^{-1}}\right) \left( {T_{\rm dust, CPD} \over 20~{\rm K}} \right) \left( {M_{\rm dust, CPD} \over {2\times10^{-5}~M_\earth}} \right).
\end{eqnarray}
With an opacity of $\sim 2~{\rm cm}^2~{\rm g}^{-1}$ for sub-$\mu$m grains at $855~\mu$m (Figure \ref{fig:opacity}) and a stellar irradiation-dominated CPD temperature of $\sim30-35$~K (Appendix \ref{sec:simobs}), Equation (\ref{eqn:cont_flux}) implies that the continuum flux is estimated to be $<1~\mu$Jy, far too small to explain the observed continuum flux of $\sim100~\mu$Jy \citep{isella19}. We may invoke in situ grain growth within CPDs; however, although a larger sub-mm grain opacity of $\sim 10~{\rm cm}^2~{\rm g}^{-1}$ helps, the estimated continuum flux is still a few $\mu$Jy, more than an order of magnitude smaller than the observed flux. 

To explain the observed continuum flux we thus need either a small CPD viscosity of $\alpha_{\rm CPD} \lesssim 10^{-5}$ or accumulation of dust in CPDs over multiple viscous timescales so that the total dust mass is much larger than the steady-state dust mass estimated. 
One possibility for the latter is that sub-$\mu$m grains grow in situ in the CPDs to sub-mm sizes and the sub-mm grains remain trapped in pressure bumps, created perhaps by existing moons and/or radially varying mass transport throughout the disk. Since sub-mm grains in CPDs have short radial drift timescales compared with the gas viscous timescale \citep[e.g.,][]{zhu18}, pressure bumps are necessary if we were to explain the continuum flux with sub-mm grains. Interestingly, this overall picture is very similar to what we infer to commonly happen in protoplanetary disks, where (sub-)$\mu$m grains are supplied from the interstellar environment, grow to millimeter sizes (and beyond), and are trapped in pressure bumps. Also, this picture is similar to some Galilean satelite formation models in the proto-Jovian disk \citep[e.g.,][]{canup02}.

We note that the above order-of-magnitude estimates are consistent with results from more detailed calculations of CPD's sub-mm continuum flux, for instance that in \citet{zhu18}. Based on Figure 3 of \citet{zhu18} we can infer that, with $M_{p}\dot{M}_{\rm acc}\sim10^{-7}~M_{\rm Jup}^2~{\rm yr}^{-1}$ and $\alpha_{\rm CPD}=10^{-3}$, the CPD continuum flux can reach 100~$\mu$Jy only when the dust-to-gas mass ratio is $f_{\rm dtg} \gtrsim 0.01$. If dust is depleted by a significant fraction at the gap edge and does not accumulate in CPDs, CPDs with $\alpha_{CPD}\sim10^{-3}$ will be much fainter than 100~$\mu$Jy. 

In our numerical simulations, test particles captured in CPDs are not accreted onto the planets. Particles hence accumulate in the CPDs as if there are pressure bumps. In the $M_c = 2.5~M_{\rm Jup}$ model, the total dust mass in the CPDs at $t=0.6$~Myr reaches $(2-3) \times10^{-3}~M_\earth$. Consistent with the estimates from Equation (\ref{eqn:cont_flux}), that is $20-30~\mu$Jy assuming sub-$\mu$m grain opacity of $\kappa_\nu = 2~{\rm cm}^2~{\rm g}^{-1}$, we find that the CPDs are too faint and are not detected at the rms noise level comparable to the observation\footnote{We find that reducing the rms noise level to $5~\mu{\rm Jy}~{\rm beam}^{-1}$ would reveal the CPDs at $3\sigma$ level.}. Note however that the observed sub-mm flux could be reproduced if sub-$\mu$m grains in the CPDs are warmer ($\gtrsim 100$~K) than the stellar irradiation-dominated temperature. If CPD's internal heating and/or planet's accretion heat play an important role, the observed sub-mm CPD flux could be explained with sub-$\mu$m grains only.

As an alternative, one may invoke in situ grain growth within the CPDs. In order to examine such a possibility, we redistribute the total CPD dust mass over a range of dust sizes from $0.1~\mu$m to 1~mm, assuming a power-law size distribution with an exponent of $-3.5$. With this implementation, most of the CPD dust mass is in sub-mm grains and the observed CPD continuum flux can be reproduced as can be seen in Figure \ref{fig:simobs}c, thanks to a larger grain opacity. Again, note that the CPDs need to have pressure bumps to trap those large grains because otherwise they are subject to rapid radial drift. 

Future higher angular resolution observations will be able to separate PDS~70c's CPD from the outer continuum ring (Figure \ref{fig:simobs}d). Because the upper layers of the near (i.e., Western) side of the outer circumstellar disk will block optical/IR emission from PDS~70c as it moves behind the upper layers, high spatial resolution observations at (sub-)mm wavelengths will be crucial if we were to constrain PDS~70c's orbit.

\subsection{Potential Implications to Planet Formation in the Solar System and Mature Exoplanetary Systems}
\label{sec:solar}

In the solar nebula, an increasing number of meteoritic isotope measurements suggests that the solar protoplanetary disk had two genetically distinct reservoirs, which coexisted and remained spatially separated \citep[e.g.,][]{warren11,kruijer17}. One possible way to explain this so-called  non-carbonaceous/carbonaceous meteorite dichotomy is if Jupiter (and possibly Saturn, too) has grown early to open a gap around its orbit within $\sim1$~Myr, preventing large particles from flowing into the inner disk \citep{kruijer17}. The sub-mm continuum ring located beyond the two forming planets in the PDS~70 disk provides a strong observational evidence that such a filtration mechanism could have operated in the solar protoplanetary disk, too. 

The Kepler mission revealed that more than $30\%$ of nearby Sun-like, FGK-type stars have at least one close-in super-Earth/mini-Neptune \citep{petigura13,zhuw18}, whereas the solar system does not have such a planet. Similar to what we see in the PDS~70 disk and our simulations, it is likely that Jupiter and Saturn reduced the inward solid mass flux in the solar nebula by trapping large grains beyond their orbits. As a result, a significantly less amount of solid would have been available in the inner disk, in which case the growth of terrestrial planets had to be limited \citep[see also e.g.,][]{haugbolle19,lambrechts19}. Future statistical comparisons of the warm/cold Jupiter occurrence rate between Earth hosting stars and super-Earth/mini-Neptune hosting stars may help reveal if giant planets indeed play a role in determining the final mass of terrestrial planets in the system.

Our simulations suggest that giant planets could settle into mean motion resonance early while they grow embedded in their gaseous host disk. In the solar system, it is proposed that Jupiter and Saturn have captured in mean motion resonance while they are embedded in the solar nebula \citep{masset01,morbidelli07,walsh11}. An early settlement into mean motion resonance is also consistent with the directly imaged four giant planets orbiting in the HR~8799 debris disk that are suggested to be in 8:4:2:1 mean motion resonance \citep{konopacky16,wang18}.

\section{CONCLUSION}
\label{sec:conclusion}

PDS~70 offers an ideal testbed for planet-disk interaction studies. Using two-dimensional hydrodynamic planet-disk interaction simulations, we show that the signposts of planet-disk interaction predicted by numerical simulations show an excellent agreement with  observed features in the PDS~70 disk. This strongly suggests that previously proposed theories of planet-disk interaction, including resonant migration, particle trapping, size segregation, and filtration, are indeed in action. In particular, the sub-mm continuum ring observed outward of the two directly imaged planets provides the first observational evidence that gap-opening planets can hold particles with appropriate sizes beyond their orbits.

By studying planets in formation and the co-evolution with their host disk, we can also infer the formation history of mature planetary systems. The fact that giant planets filter large grains suggests gas giants can have an influence on the formation and characteristics of terrestrial planets in the system, as they can induce a chemical inhomogeneity and reduce the inward solid mass flux. It will be interesting to know if it is common for giant planets to settle in a mean motion resonance while they are growing embedded in the host disk. Transitional disks having inner cavities might be good targets for this purpose.

Although we show that two already-grown giant planets placed close to 2:1 mean motion resonance can quickly settle into an orbital resonance, our simulations do not address how the two planets have reached such a configuration in the first place. On one hand, it is possible that the two planets form and grow near the commensurability. In this case, the two planets should have grown at a rate of $\gtrsim 10^{-6}~M_{\rm Jup}~{\rm yr}^{-1}$ on average, potentially punctuated by even higher accretion rates during episodic accretion events. Whether or not the planets can remain dynamically stable with such mass growth rates have to be examined. On the other hand, the two planets could have formed and grown at distance, but later migrated toward each other and captured in a mean motion resonance. In this case, the migration must be sufficiently slow so that the planets do not cross the 2:1 mean motion resonance but also they do not trigger a dynamical instability. Future simulations taking into account both orbital migration and mass growth from one to a few tens Earth-mass cores will help infer the full history of PDS~70b and c's formation and evolution.

\acknowledgments

We thank the anonymous referee for prompt and helpful reports.
This paper makes use of the following ALMA data: ADS/JAO.ALMA \#2015.1.00888.S,   ADS/JAO.ALMA \#2017.A.00006.S. ALMA is a partnership of ESO (representing its member states), NSF (USA) and NINS (Japan), together with NRC (Canada), MOST and ASIAA (Taiwan), and KASI (Republic of Korea), in cooperation with the Republic of Chile. The Joint ALMA Observatory is operated by ESO, AUI/NRAO and NAOJ. The National Radio Astronomy Observatory is a facility of the National Science Foundation operated under cooperative agreement by Associated Universities, Inc. JB acknowledges support by NASA through the NASA Hubble Fellowship grant \#HST-HF2-51427.001-A awarded  by  the  Space  Telescope  Science  Institute,  which  is  operated  by  the  Association  of  Universities  for  Research  in  Astronomy, Incorporated, under NASA contract NAS5-26555. JB acknowledges support from NASA grant NNX17AE31G and computing resources provided by the NASA High-End Computing (HEC) Program through the NASA Advanced Supercomputing (NAS) Division at Ames Research Center. ZZ acknowledges support from the National Aeronautics and Space Administration through the Astrophysics Theory Program with grant No. NNX17AK40G and support from the Sloan Research Fellowship. LP acknowledges support from CONICYT project Basal AFB-170002 and from FONDECYT Iniciaci\'on project \#11181068. RT acknowledges support from the Smithsonian Institution as a Submillimeter Array (SMA) Fellow.

\software{\texttt{Dusty FARGO-ADSG} \citep{baruteau19}, \texttt{RADMC-3D} \citep{radmc3d}}

\appendix

\section{Initial Disk Temperature}
\label{sec:disk_temperature}

To set up the initial disk temperature profile, we first construct a stellar irradiation-dominated temperature profile $T_{\rm irr}$: 
\be
\label{eqn:t_irr}
T_{\rm irr}(R) = \left( {f L_* \over 4\pi R^2\sigma_{\rm SB}} \right)^{1/4}.
\en
Here, $f=0.1$ is introduced to account for the non-normal irradiation at the disk
surface, $L_*= 0.35$~\lsun \citep{pecaut16,keppler18} is the stellar luminosity, and $\sigma_{\rm SB}$ is the Stefan-Boltzmann constant.
Assuming that the disk has a constant temperature over height at each radius, we construct the three-dimensional gas density structure that satisfies the hydrostatic equilibrium in the vertical direction
\be
- {GM_* Z \over (R^2 + Z^2)^{3/2}} - {1 \over \rho}{\partial P \over \partial Z} = 0.
\en
Here, $G$ is the gravitational constant, $M_*$ is the stellar mass, $Z$ is the height, $\rho$ is the three-dimensional gas density, and $P=\rho \mathcal{R} T/\mu$ is the gas pressure with $\mathcal{R}$ and $\mu$ being the gas constant and mean molecular weight. We adopt the meridional boundary of the three-dimensional MCRT calculation domain at $Z/R=\pm0.45$ (i.e., $\pm24$ degrees against the midplane). At the location of the two planets, this covers more than 6 scale heights with the stellar irradiation-dominated disk temperature. 

\begin{figure*}[b!]
    \centering
    \includegraphics[width=1\textwidth]{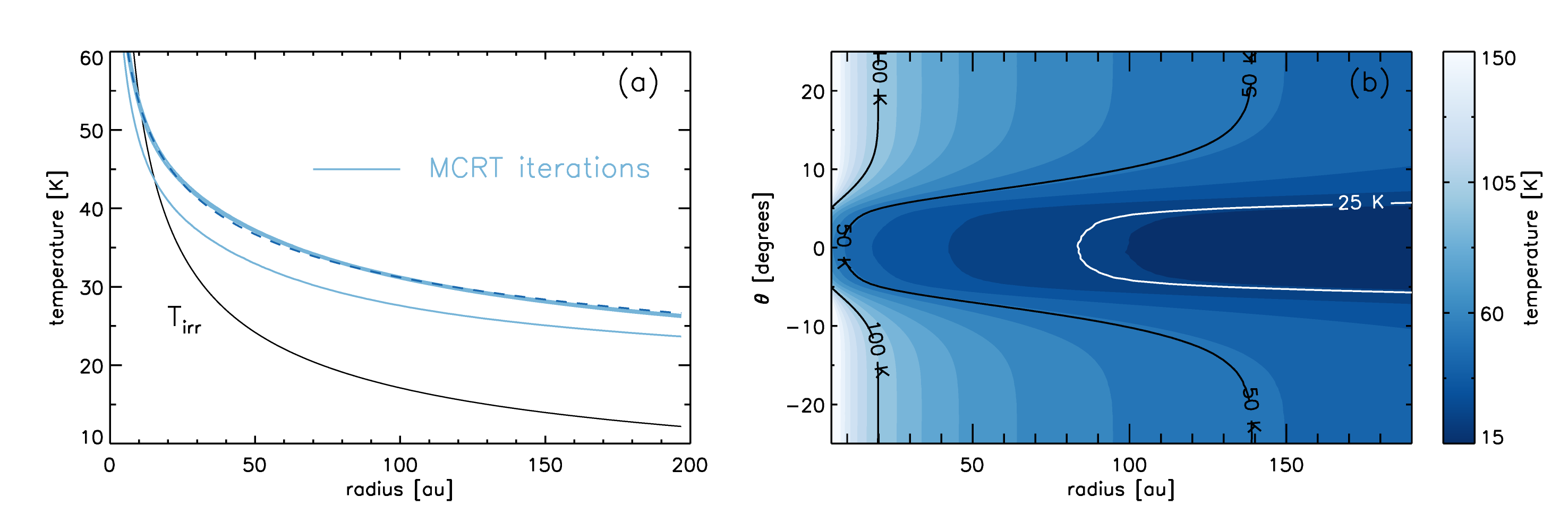}
    \caption{(a) Blue curves show the density-weighted, vertically integrated temperature distribution $\bar{T}$ from the first 10 MCRT iterations. The blue dashed curve shows the power-law fit to the temperature profile obtained from the 10th iteration: $\bar{T} (R) = 44~{\rm K}~(R/22~{\rm au})^{-0.24}$. The black curve shows the stellar irradiation-dominated temperature $T_{\rm irr}$ (Equation \ref{eqn:t_irr}). (b) The two-dimensional $R-Z$ temperature distribution after 10 MCRT iterations.}
    \label{fig:temp1}
\end{figure*}

We fix the gas surface density profile, as described in Equation (\ref{eqn:sigma}), and iterate MCRT calculations to compute the three-dimensional disk temperature profile using RADMC-3D \citep{radmc3d}. After each MCRT calculation, the three-dimensional gas density structure is updated with the new temperature. We stop the iteration when the density-weighted, vertically integrated temperature distribution $\bar{T}$, defined as
\be
\bar{T}(R) = {{\int{\rho(R,Z)T(R,Z) dZ}} \over {\int{\rho(R,Z) dZ}}},
\en
does not vary more than $1\%$ at each radius from the previous iteration. We find that $\bar{T}$ converges within the first few iterations (Figure \ref{fig:temp1}) and the final temperature profile is not sensitive to the choice of the input temperature profile. 

In each Monte Carlo calculation we use $10^9$ photon packages. We assume total $2\times10^{-7}~M_\odot$ of small dust between 0.01 and 0.1~$\mu$m with a power-law size distribution, adopting a power-law exponent of $-3.5$. We assume that small grains are perfectly coupled with disk gas. We assume these small grains are compact monomers consist of $60~\%$ silicate and $40~\%$ amorphous carbon, having an internal density of $2.7~{\rm g~cm}^{-3}$. We adopt optical constants of silicate and amorphous carbon from \citet{draine84} and \citet{li97}, respectively.

\section{Simulated Continuum Observation}
\label{sec:simobs}

To produce the simulated continuum images presented in Figure \ref{fig:simobs}, we take the gas density distribution at $t=0.6$~Myr and run MCRT iterations as explained in Appendix \ref{sec:disk_temperature}. One additional step we had is that, before we expand the two-dimensional surface density from the hydrodynamic simulation in the vertical direction, we remove the material within the Hill sphere of the planets and fill the region with the azimuthally averaged circumstellar disk surface density at the corresponding radial location. This is necessary because the CPDs have their own scale heights which are much smaller than that of the circumstellar disk. The temperatures at the location of the CPDs from the MCRT iterations are therefore stellar irradiation-dominated temperature. If internal (e.g., viscous) and/or planet's accretion heat play a role, the CPD temperature can be higher than the stellar irradiation-dominated temperature.

\begin{figure*}[h!]
    \centering
    \includegraphics[width=1\textwidth]{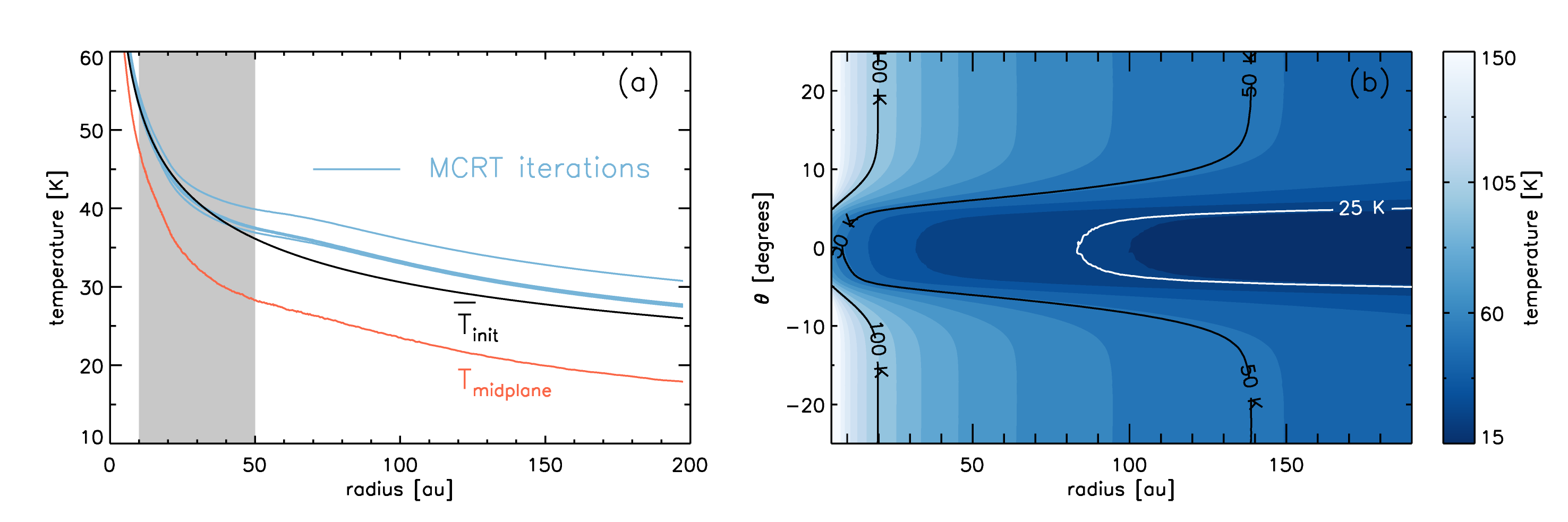}
    \caption{Similar to Figure \ref{fig:temp1}, but with the gas density distribution at $t=0.6$~Myr in the $M_c=2.5~M_{\rm Jup}$ model. In panel (a), the shaded region shows the radial location of the common gap opened by PDS~70b and c. The orange curve shows the disk midplane temperature obtained after 10 MCRT iterations. The black curve shows the initial $\bar{T} (R)$ distribution (Equation \ref{eqn:temperature}).}
    \label{fig:temp2}
\end{figure*}

The temperature profile from MCRT iterations is shown in Figure \ref{fig:temp2}. The temperature is slightly lower than the initial temperature within the gap and is higher beyond the gap because the outer disk is more directly exposed to the stellar photons.

\begin{figure*}[h!]
    \centering
    \includegraphics[width=1\textwidth]{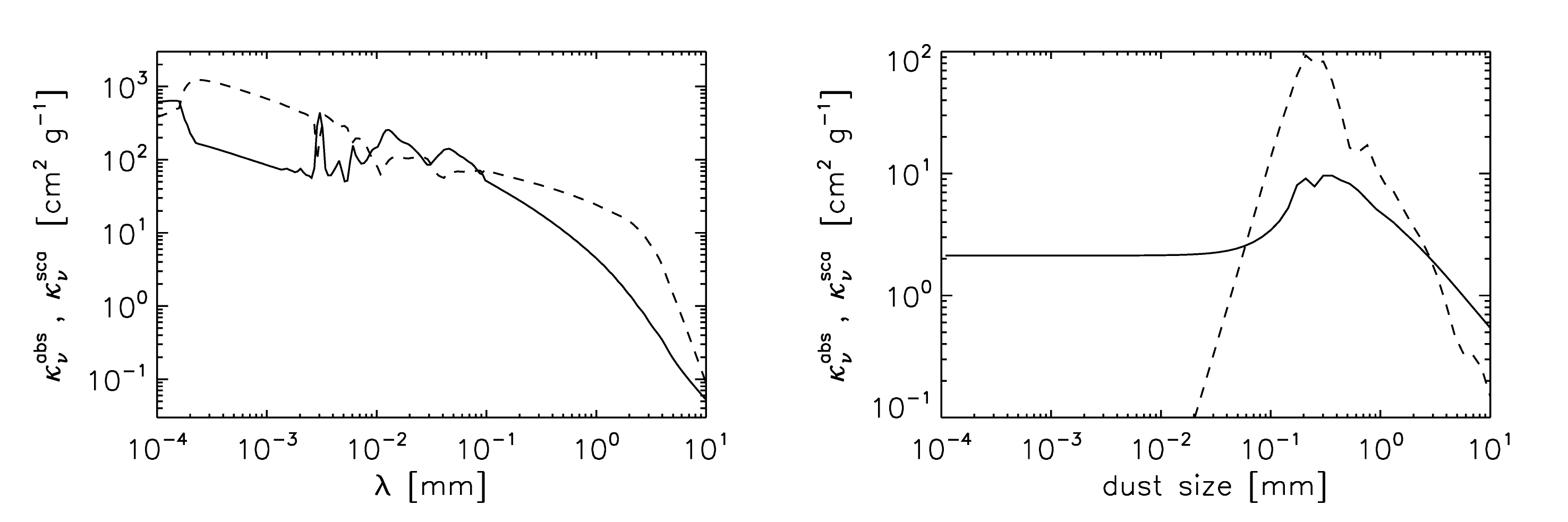}
    \caption{(Left) The absorption (solid) and scattering (dashed) opacities $\kappa_\nu^{\rm abs}$ and $\kappa_\nu^{\rm sca}$ as a function of the observing wavelength $\lambda$. (Right) The absorption (solid) and  scattering (dashed) opacities at $\lambda = 855~\mu$m, as a function of dust size.}
    \label{fig:opacity}
\end{figure*}

We assume that grains are composed of $30~\%$ silicate matrix and $70~\%$ water ice, having an internal density of $1.26~{\rm g~cm}^{-3}$. The optical constants of water ices and astrosilicates are adopted from Jena database and \citet{draine84}, respectively. In the left panel of Figure \ref{fig:opacity}, we show the size averaged absorption opacity as a function of the observing wavelength, assuming a power-law size distribution with a power-law exponent of $-3.5$ and the minimum and maximum grain size of $0.1~\mu$m and 1~mm. Presented in the right panel is size-dependent absorption and scattering opacities at $\lambda=855~\mu$m. Simulated continuum images are produced considering both absorption and anisotropic scattering using the Henyey-Greenstein approximation. 

%\bibliography{bibliography}{}
%\bibliographystyle{aasjournal}

\end{document}